\documentclass[pra,twocolumn,showpacs,superscriptaddress,nofootinbib]{revtex4-1}
\usepackage{graphicx}
\usepackage{bm}
\usepackage{amssymb}
\usepackage{amsmath}
\usepackage{amstext}
\usepackage{latexsym}
\usepackage[colorlinks,linkcolor=blue]{hyperref}
\usepackage{bm}
\def\tr{\rm{Tr}}
\def\la{{\langle}}
\def\ra{{\rangle}}

\def\e{\epsilon}
\def\q{\quad}
\def\d{{\rm d}}

\def\t{\tilde{t}}

\def\h{\hat{H}}

\begin{document}
\title{Fidelity of fermionic atom-number states subjected to tunneling decay}
\author{M. Pons}
\affiliation{Departmento de F\' isica Aplicada I, EUITMOP, Universidad del Pa\' is Vasco, UPV-EHU, Barakaldo, pain}
\author{D. Sokolovski}
\affiliation{Departmento de Qu\' imica-F\' isica, Universidad del Pa\' is Vasco, UPV-EHU, Leioa, Spain}
\affiliation{IKERBASQUE, Basque Foundation for Science,
E-48011 Bilbao, Spain}
\author{A. del Campo}
\affiliation{Institut f{\"u}r Theoretische Physik, Leibniz Universit\"at Hannover, 
Hannover, Germany}

\def\G{\Gamma}
\def\L{\Lambda}
\def\la{\lambda}
\def\g{\gamma}
\def\al{\alpha}
\def\s{\sigma}
\def\e{\epsilon}
\def\k{\kappa}
\def\ve{\varepsilon}
\def\l{\left}
\def\r{\right}
\def\te{\mbox{e}}
\def\d{{\rm d}}
\def\t{{\rm t}}
\def\K{{\rm K}}
\def\N{{\rm N}}
\def\H{{\rm H}}
\def\la{\langle}
\def\ra{\rangle}
\def\om{\omega}
\def\Om{\Omega}
\def\vep{\varepsilon}
\def\wh{\widehat}
\def\tr{{\rm tr}}
\def\da{\dagger}
\def\iz{\left}
\def\zi{\right}
\newcommand{\beq}{\begin{equation}}
\newcommand{\eeq}{\end{equation}}
\newcommand{\beqa}{\begin{eqnarray}}
\newcommand{\eeqa}{\end{eqnarray}}
\newcommand{\intf}{\int_{-\infty}^\infty}
\newcommand{\into}{\int_0^\infty}

\begin{abstract}
Atom-number states are a valuable resource for ultracold chemistry, atom interferometry and quantum information processing. Recent experiments have achieved their deterministic preparation in trapped few-fermion systems.
We analyze the tunneling decay of these states, both in terms of the survival probability,
and the non-escape probability, which can be extracted from measurements of the full counting statistics.
At short times, the survival probability  exhibits deviations from the exponential law. The decay is governed by the multi-particle Zeno time which exhibits a signature of quantum statistics and contact interactions. The subsequent exponential regime governs most of the dynamics, and we provide
accurate analytical expressions for the associated decay rates.
Both dynamical regimes are illustrated in a realistic model. Finally, a global picture of multi-particle quantum decay is presented.

\end{abstract}

\maketitle
\section{Introduction}
\begin{figure}
\includegraphics[width=0.9\linewidth]{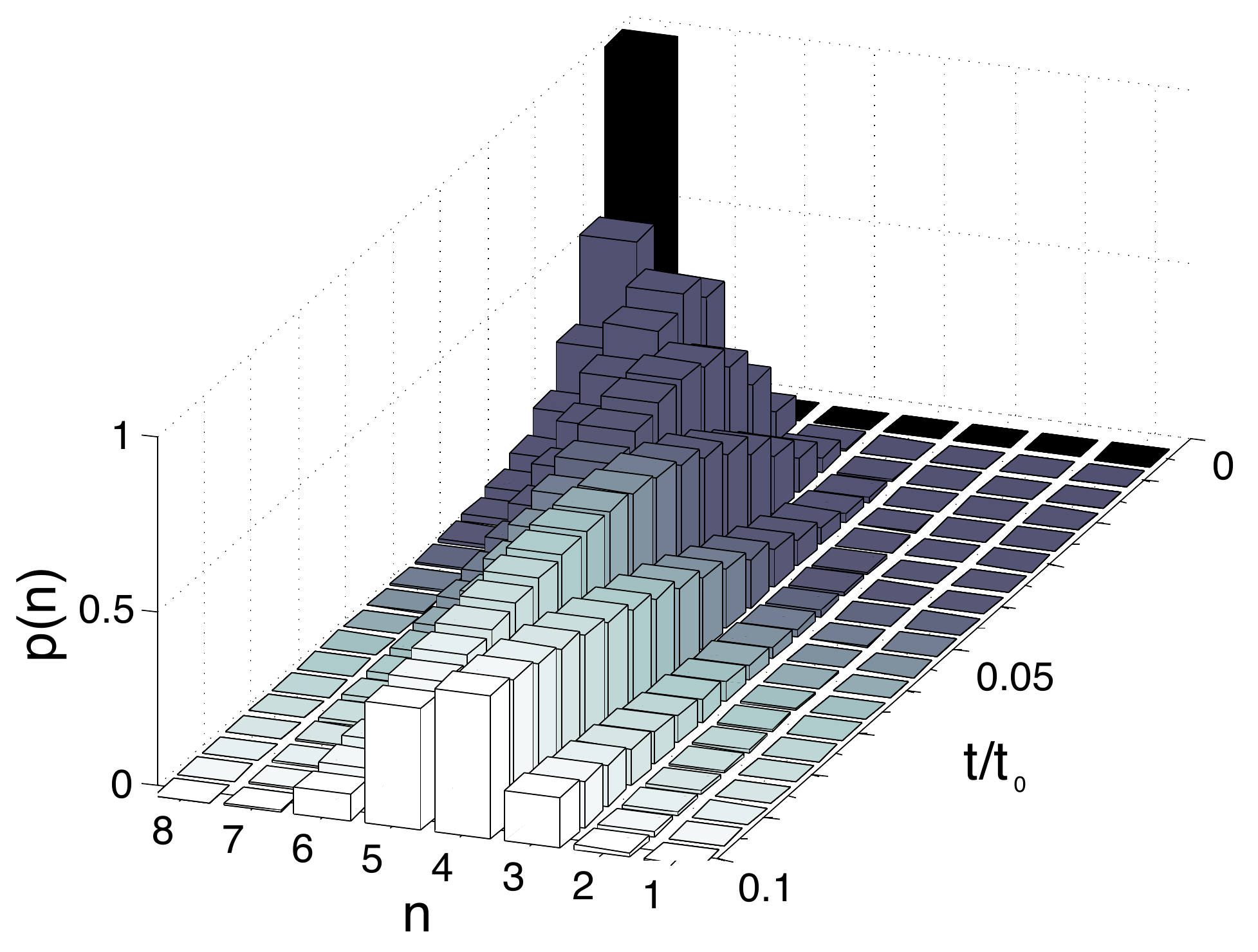}
  \caption{
\label{pnt}
Fidelity decay of an atom-number state of polarized fermions induced by tunneling losses.
The probability to find $n$ atoms inside the trap, $p(n,t)$, is initially peaked at $n=8$. As the time of evolution goes by, $p(n,t)$ broadens and shifts to lower values of the mean atom number.
The trapping potential model is further discussed in Sect.\ref{secmodel}.}
\end{figure}

Successful production and preservation of atomic states containing an exactly known number of particles  (so-called atomic Fock or atom-number states) is of importance
 for ultracold chemistry, atom interferometry, quantum information processing and investigating the foundations of quantum physics.
Among the methods proposed for creating  atom-number states are {\it atom culling} in time dependent traps
\cite{DRN07,MUG,PONS,WRN09,RWZN09,DIMA} and the  use of similar techniques in optical lattices  \cite{Qdistill, NP10}. Spectacular experimental progress in the deterministic preparation of few-atom states has been reported in
 \cite{exp1,exp2}.
But when the confining potential has finite potential barriers, which allow tunneling leakage,  an initial trapped state with a well-defined atom number eventually evolves into a mixture of several atom-number states, as shown in Fig. \ref{pnt}.
This process can be quantitatively characterized by the fidelity decay. The fidelity is often defined as the probability to persist in the initial atom-number state or, somewhat less stringently, by the non-escape probability, i.e., the probability that the total number of atoms within the trap remains the same. Unlike the integrated density profile of an atomic cloud  \cite{DDGCMR06,Ceder09}, both probabilities
refer to {\it all} particles occupying a certain subspace of a Hilbert space simultaneously and are, in this sense, multi-particle observables.
%

The purpose of this paper is to investigate the fidelity decay of an atomic Fock state due to tunneling leakage. Hence, it is both of fundamental interest to understand the quantum decay of multi-particle systems, as well as of practical relevance to determine the decay rates of trapped
Fock states \cite{exp1,exp2,DRN07,MUG,PONS,WRN09,RWZN09,DIMA}.
We focus on ultracold atomic vapors confined in tight, effectively one-dimensional,  waveguides and, more specifically, on polarized fermions and related fermionized systems such as a bosonic cloud in the Tonks-Girardeau regime \cite{Girardeau60}.
Since the rate of atom losses due to three-body collisions is suppressed due to spatial anti-bunching \cite{g3bosons,g3fermions}, these  systems are well suited for investigating the quantum  dynamics of multi-particle tunneling decay. Although the tunneling decay is expected to roughly follow an exponential law, deviations are expected both at short and long times on theoretical grounds.
Long-time deviations from exponential decay in the tunneling dynamics of multi-particle systems  have been studied in \cite{delcampo11}, and we shall focus on the short-time and exponential regime.
Additional relevant works include studies of the fidelity decay in a multi-particle Loschmidt echo \cite{Goold11,Buljan11} and different scenarios of two-particle quantum decay \cite{MG10,TS11,GCML11,KB11,KLW11}.
The rest of the paper is organized as follows: in Section \ref{SecNEP} we define the non-escape and survival probabilities for a general quantum system. In Section \ref{shorttimes} we analyze their short-time evolution and the Zeno effect \cite{Z1,Z2,shortexp}. In Section \ref{Zenofermions} we discuss the decay of one-dimensional trapped fermions and related systems and investigate the effect of quantum statistics in terms of the multi-particle Zeno time.
Section \ref{expdecay} is devoted to the regime governed by an exponential decay law, an accurate semiclassical approximation for the corresponding decay rates is provided. Section \ref{secmodel} illustrates the different dynamical regimes in a trap model of relevance to recent experiments \cite{exp1,exp2}. A general picture of multi-particle quantum decay is outlined in Section \ref{discuss}, before closing the paper with a set of conclusions in Section \ref{conclu}.

\section{Non-escape and survival probabilities \label{SecNEP}}

We start by considering the time evolution of the probability that a multi-particle quantum system, initially prepared in a (possibly mixed) state $\varrho_0$, will be found inside a certain sub-space
$\mathcal{D}\subset\mathcal{H}$ of the Hilbert space $\mathcal{H}$.
The ability to experimentally measure the Full Counting Statistics (FCS) \cite{exp1} motivates the introduction of the multi-particle non-escape probability \cite{delcampo11}. Consider the full atom-number distribution
$p(n,t)=\la \delta(\hat{n}_{\mathcal{D}}-n)\ra_t$ where $\hat{n}_{\mathcal{D}}$ is the atom-number operator in the subspace of interest $\mathcal{D}$, $n=1,\dots,\N$ is an integer, and the expectation value is taken with respect to a time evolving state. $p(n,t)$ represents the probability of finding exactly $n$-particles at time $t$ in the subspace $\mathcal{D}$. Its typical behavior is exhibited in Fig. \ref{pnt} for a $\N$-particle metastable Fock state.
The probability to preserve the $N$-particle Fock state at time $t$ is given by  $p(n=\N,t)$ and we shall denote it by $P_{\mathcal{D}}(t)$ in the following.

Letting $\hat{\Pi}$ be the projector on $\mathcal{D}$, we write this probability as
 \begin{eqnarray}\label{2}
\mathcal{P}_{\mathcal{D}}(t)=\tr[\varrho_0 \hat{\Pi}(t)], \quad{\hat{\Pi}(t)}=\exp(i\h t)\hat{\Pi}\exp(-i\h t).\q
\end{eqnarray}
Let $|n\ra$ be a basis on $\mathcal{H}$, so that an initial state $\varrho_0=\sum_n p_n|n\ra\la n|$.
Let $\hat{\Lambda}=\sum_n|n\ra\la n|$ be the projector associated with the space spanned by the initial state $\varrho_0$, such that $\hat{\Lambda}\varrho_0\hat{\Lambda}=\varrho_0$.
The behavior of $P_{\mathcal{D}}(t)$ largely depends on whether $\varrho_0$ is contained within $\mathcal{D}$.
If it is, $\hat{\Pi}$ can be decomposed as the sum of projectors
 \begin{eqnarray}\label{3}
 \hat{\Pi}=\hat{\Lambda}+\hat{Q}, \q \hat{Q}\varrho_0\hat{Q} =0.
\end{eqnarray}
In this case we will refer to $\mathcal{P}_{\mathcal{D}}(t)$,  satisfying $\mathcal{P}_{\mathcal{D}}(0)=1$, as the  {\it non-escape probability} probability ${\rm P}(t)$ which must be a non-increasing function at least for short times. In the special case
of $\hat{Q}=0$, $P_{\mathcal{D}}(t)$ becomes the {\it survival  probability}, i.e., that for the system to remain in its initial state. This we will denote
 by ${\rm S}(t)$, bearing in mind that ${\rm S}(0)=1$. Note that ${\rm P}(t) \ge {\rm S}(t)$ since the condition for the system to stay within $\mathcal{D}$,
is less strict than that for
staying in  one specific state $\varrho_0$. An obvious example is a wavepacket contained within a spatial
region $\Delta$, for which the non-escape condition means not leaving $\Delta$, as opposed to remaining in the initial state (survival).
Finally, if $\varrho_0$ is not contained in $\mathcal{D}$ we have
  \begin{eqnarray}\label{4}
\varrho_0= \hat{\Pi}\varrho_0\hat{\Pi}+ \delta \varrho_0
\end{eqnarray}
where $\delta \varrho_0$ is the component of the initial
state orthogonal to $\mathcal{D}$.
In this case $P_{\mathcal{D}}(t)$ may increase even at short times, a simple example
being a wavepacket arriving in an initially empty region of space. Next we consider the short time
evolution of the two probabilities.

\section{The short time limit\label{shorttimes}}

The short time expansion of the probability to remain in the sub-space $\mathcal{D}$ takes the form
\beqa\label{1a}
\mathcal{P}_{\mathcal{D}}(t)&=&
\tr(\varrho_0\hat{\Pi})-i\tr(\varrho_0[\hat{\Pi},\hat{H}])t\nonumber\\
& &-\tr[\varrho_0(\frac{1}{2}\{\hat{\Pi},\hat{H^2}\}-\hat{H}\hat{\Pi}\hat{H})]t^2+\mathcal{O}(t^3), \nonumber\\
\eeqa
where $\{\hat{A},\hat{B}\}$ denotes the anticommutator of $\hat{A}$ and $\hat{B}$.
As it is well known (see, e.g.,  \cite{Z1,Z2}),  vanishing of the term linear in $t$ leads to the Zeno effect:
frequent checks of whether the system is still contained in ${\mathcal{D}}$ (e.g., projection measurements of $\hat{\Pi}$) would prevent it from leaving the sub-space. In particular, with the interval between checks being $\tau$, $\mathcal{P}_{\mathcal{D}}(t)$ would typically exhibit an effective exponential decay
 \begin{eqnarray}\label{4aa}
\mathcal{P}_{\mathcal{D}}(t)\sim \exp(-\gamma t), \q \gamma\equiv \tau/\tau_Z^2,
\end{eqnarray}
where the Zeno time $\tau_Z$, is determined by the coefficient multiplying $t^2$ in Eq. (\ref{1a}),
 \begin{eqnarray}\label{2a}
\tau_Z=[\tr[\varrho_0(\{\hat{\Pi},\hat{H^2}\}/2-\hat{H}\hat{\Pi}\hat{H})]^{-1/2}.
\end{eqnarray}
Nonetheless, we shall be concerned with quantum decay in the absence of measurements.
It is readily seen that the linear in time term in Eq. (\ref{1a}) vanishes provided the initial state is contained in $\mathcal{D}$, i.e. the orthogonal component $\delta \varrho_0=0$, or  can be neglected, $||\delta \varrho_0||\ll 1$. The decay of $\mathcal{P}_{\mathcal{D}}(t)$ becomes then quadratic in time
and governed by $\tau_Z$ according to
\beqa
\mathcal{P}_{\mathcal{D}}(t)=1-(t/\tau_Z)^2+\mathcal{O}(t^3).
\eeqa
Under this condition, from Eq. (\ref{2a}) for the Zeno time, we have
 \begin{eqnarray}
\label{4a}
\tau_Z= [\Delta\hat{H}_{\varrho_0}-\tr(\varrho_0\hat{H} \hat{Q}\hat{H})]^{-1/2},
\end{eqnarray}
and not merely given by the inverse of the energy variance of the initial state,
$\Delta\hat{H}_{\varrho_0}=\tr[\varrho_0\hat{H^2}]-\tr[\varrho_0\hat{H}]^2$.
The last term
in the square bracket vanishes in the case of the survival probability,
$ \hat{Q}=0$. Thus, a survival Zeno time never exceeds that in the non-escape case, reflecting
the fact that it is easier to maintain a system within a larger subset of its Hilbert space.
(Note that $\tau_Z$ in Eq. (\ref{4a}) becomes infinite, as it should, if $\mathcal{D}$ is chosen to coincide with $\mathcal{H}$). Next we proceed to the case of several particles confined in a potential trap with tunneling leakage.

\section{Multi-particle Zeno times of fermionized systems \label{Zenofermions}}

Consider $\N$ non-interacting particles initially trapped in a potential well $V_I(x)$  which,
at $t=0$, is instantly converted into a trapping potential with a finite barrier, $V(x)$, as that in Fig. \ref{fcs} below, thus allowing the particles to escape into the continuum.
The Hamiltonian of the system reads
 \begin{eqnarray}\label{1b}
\hat{H}(x_1,...,x_\N)&=&\sum_{i=1}^\N\hat{h}(x_i),\nonumber\\
\hat{h}(x_i)&=&-\partial_{x_i}^2/2+V(x_i), \q i=1,2,...,\N.
\end{eqnarray}
Spin polarized fermions fall within this description, and certain strongly interacting systems can be described in a similar way. This is the case of bosonic atoms in the Tonks-Girardeau regime, where strong zero-range hard-core interaction leads to fermionization \cite{Girardeau60}.
In this case there is a one-to-one correspondence (the Bose-Fermi mapping \cite{Girardeau60}) between the symmetric state of strongly interacting bosons,  $\Psi^{TG}(x_1,\dots,x_\N)$, and the anti-symmetric state of the dual system of non-interacting fermions, $\Psi^F(x_1,\dots,x_\N)$,
\beqa
\Psi_{TG}(x_{1},\dots,x_{\N})= \mathcal{A}\Psi_{F}(x_{1},\dots,x_{\N}),
\eeqa
where the antisymmetric unit function $\mathcal{A}=\prod_{1\leq j<k\leq \N}{\rm sgn}(x_{k}-x_{j})$.
It follows that for the Tonks-Girardeau gas both the non-escape and the survival probabilities coincide with those calculated for the corresponding system of non-interacting fermions \cite{delcampo11},
\beqa
\label{2bc}
\mathrm{P}^{TG}(t)=\mathrm{P}^{F}(t),\q \mathrm{S}^{TG}(t)=\mathrm{S}^{F}(t).
\eeqa
Both of these systems were found to be optimal for the preparation of atomic Fock States using atom culling techniques \cite{exp1,exp2,DRN07,MUG,PONS,WRN09,RWZN09,DIMA}.
In the following discussion we will simply refer to a ``fermionic'' system, bearing in mind that
the results apply both to the Tonks-Girardeau gas with infinitely strong contact interactions and to a system of noninteracting polarized fermions.

The ground state of a fermionic system is given by the Slater determinant, and we choose
\beqa
\label{2b}
\Psi_0^{F}(x_{1},\dots,x_{\N}) =\frac{1}{\sqrt{\N!}}{\rm det}_{n,k=1}^{\N}[\phi_{n}(x_{k})],
\eeqa
where $|\phi_{n}\ra$, $n=1,...,\N$, are the $\N$ lowest eigenstates of the one-particle Hamiltonian,
$\hat{h}|\phi_{n}\ra=\epsilon_n|\phi_{n}\ra$.
In order to study the effect of quantum statistics on the Zeno time it is instructive to consider also the ground state for distinguishable particles,
\beqa
\label{3b}
\Psi_0^{dist}(x_{1},\dots,x_{\N}) =\prod_{n=1}^\N\phi_{n}(x_{n}),
\eeqa
and the somewhat artificial case of a bosonic excited state where the same first $\N$ levels are occupied by non-interacting particles obeying Bose-Einstein statistics,
\beqa
\label{4b}
\Psi_0^{B}(x_{1},\dots,x_{\N}) =\frac{1}{\sqrt{\N!}}{\rm per}_{n,k=1}^{\N}[\phi_{n}(x_{k})],
\eeqa
where ${\rm per}$ stands for the permanent, i.e., the sum in the r.h.s. of all
permutations of indices $k$ for a fixed order of indices $n$.
Note that these three states, Eqs. (\ref{2b}), (\ref{3b}),  and (\ref{4b}), have all the same energy.

The survival Zeno times [Eq. (\ref{4a}) with $\hat{Q}=0$] for the three cases are calculated
by inserting in Eq. (\ref{4a}) the appropriate initial state, (\ref{2b}), (\ref{3b}) or (\ref{4b}).
For distinguishable particles the calculation is straightforward, for polarized fermions
the matrix elements in Eq. (\ref{4a}) can be evaluated using the Slater-Condon rules \cite{SL}, and for bosons an extension of the later is required to symmetric states.
The result can be written in a compact form,
\beqa
\label{5b}
\tau_Z=
\bigg\{\sum_{n}\Delta\hat{h}_n +2\alpha\sum_{n<k} |\la\phi_{n}|\hat{h}|\phi_{k}\ra|^2\bigg\}^{-\frac{1}{2}},
\eeqa
where $\Delta\hat{h}_n=\la\phi_{n}|\hat{h}^2|\phi_{n}\ra
-\la\phi_{n}|\hat{h}|\phi_{n}\ra^2$, and $\alpha= 0$ for distinguishable particles, $\alpha= -1$ for fermions, and
$\alpha=1$ for excited bosons.

In Eq. (\ref{5b}), the first sum, common to all statistics, depends on the spread of the energy in the initial one-particle states induced by the tunneling decay. This contribution dominates the Zeno time. The second sum, arising from the indistinguishability of the particles and the associated symmetrization of the initial state, and more precisely, from its immanant (either determinant or permanent) structure. While it reduces the survival Zeno time for non-interacting excited bosons, it increases $\tau_Z$ in polarized fermions and bosons in the TG regime, leading to a slowing down of their decay. The symmetrization imposed by exchange quantum  statistics plays no role, as it follows from the fact that the multi-particle Zeno time is shared by dual systems related by the Bose-Fermi mapping. The different corrections to $\tau_Z$ arise only for indistinguishable particles, and are manisfest thanks to the effect that contact interactions (including as such the Pauli exclusion principle) have on the energy dispersion of the initial state.
We recall that short time decay $\mathcal{P}_{\mathcal{D}}(t)=1-(t/\tau_Z)^2=\mathcal{O}(t^3)$ is governed by the Zeno time. Also, we recall that when the system is frequently observed its lifetime $\gamma^{-1}$ becomes proportional to $\tau_Z^2$, see Eq. (\ref{4aa}).
It follows that  a fermionic state (or bosons in the TG regime) decays at short-times more slowly than non-interacting excited bosons. This result is somewhat counter intuitive in nature, given that as a result of the Pauli exclusion principle (or hard-core contact interactions in the TG gas) fermionized systems exhibit spatial anti-bunching, while non-interacting bosons prefer to group together (bunching).
However, this intuition applies only in subsequent stages of evolution, while the short-time dynamics is exclusively governed by the energy dispersion of the initial state.

In the following section we shall describe the rates associated with the subsequent exponential decay,
where the indistinguishability of the particles and density-density correlations play
a subdominant role with respect to the energy distribution.
The intuition based on spatial bunching applies to  the long-time asymptotics of multi-particle quantum decay.  At long times, deviations from the exponential decay occur due to the possibility that the decay products recombine to reconstruct the initial state. As a result, spatial bunching and antibunching effects play the dominant role, and fermionized systems decay according to a power law $\mathcal{P}_{\mathcal{D}}(t)\propto 1/t^{\alpha}$ with an exponent $\alpha>0$ about $N$ times larger than in the case of non-interacting bosons \cite{delcampo11}.

\section{Exponential regime and decay rates of fermionized systems\label{expdecay}}

\begin{figure}
\includegraphics[width=0.6\linewidth]{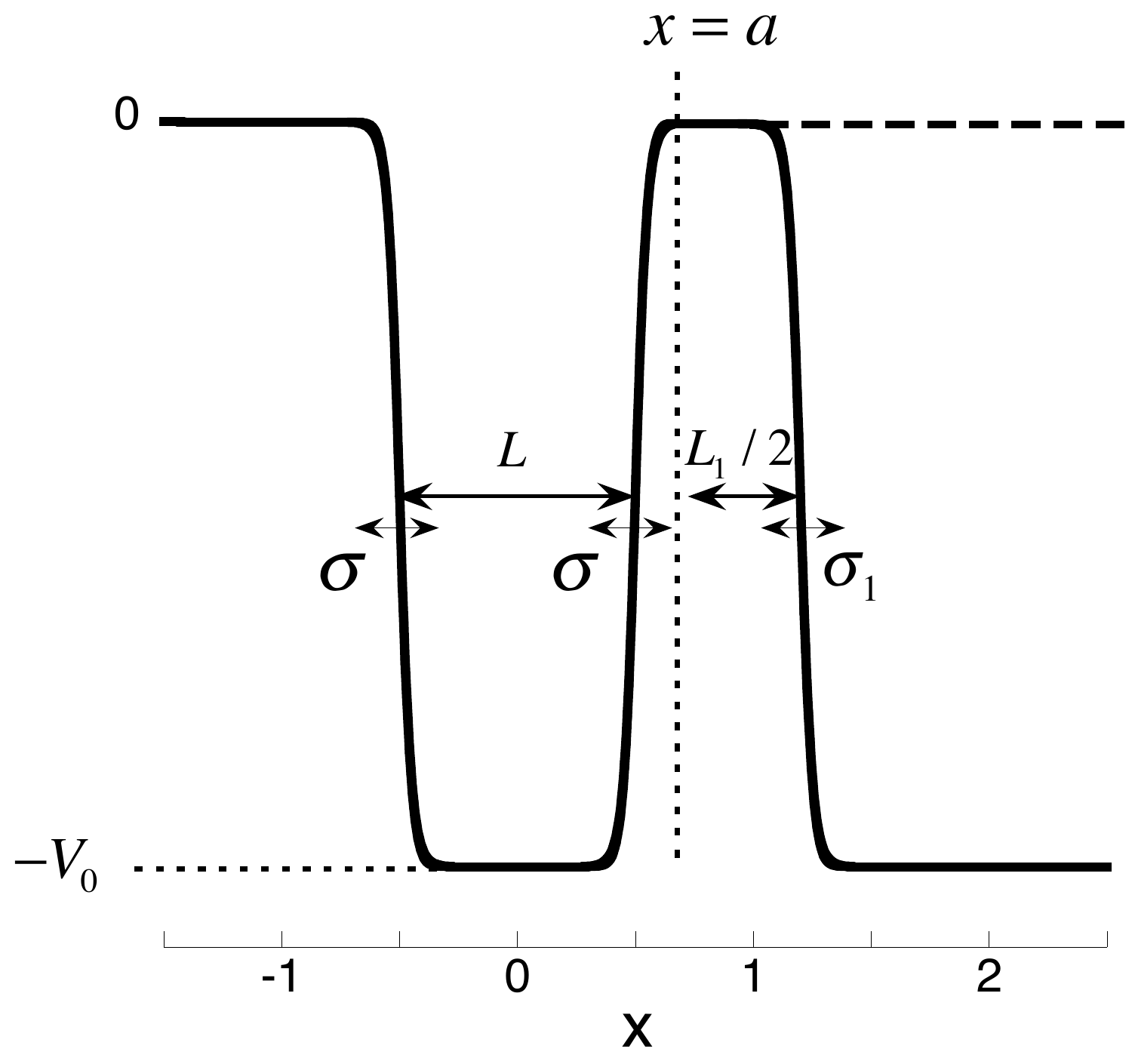}
  \caption{
\label{fcs}
The `bathtub' trapping potential $V(x)$ (solid line). Also shown is the initial well $V_I(x)$ (dashed line)}
\end{figure}

We shall now turn our attention to the regime characterized by exponential decay, most easily observed in experiments.
While manipulating cold atoms one may be interested in maintaining exactly $\N$ of them in a specified region of space $\Delta$, usually large enough as to enclose the trap subspace. The corresponding non-escape probability is obtained from Eq. (\ref{2}) by choosing the projector onto the corresponding part of the multi-configurational space,
 \begin{eqnarray}
\label{1c}
 \hat{\Pi} = \prod_{n=1}^\N \int_{\Delta} |x_n\ra \la x_n|\d x_n.
\end{eqnarray}
For a fermionic system starting in its ground state (\ref{2b}) the non-escape probability reads \cite{delcampo11}
(we drop the superscript $F$ in the following and include a subscript for the particle number $\N$)
\beqa
\label{2c}
\mathrm{P}_{\rm N}
(t)\!={\rm det}_{n,k=1}^{\N} [\la\phi_{n}(t)|\chi_{\Delta}|\phi_{k}(t)\ra],
\eeqa
 where $\la\phi_{n}(t)|\chi_{\Delta}|\phi_{k}(t)\ra=\int_{\Delta}\d x \phi_n^*(x,t)\phi_k(x,t)$,
while for the probability to survive in the state (\ref{2b}), $S_\N(t)$, one has \cite{Goold11,Buljan11,delcampo11}
\beqa
\label{3c}
\mathrm{S}_{\rm N}
(t)
&=&
\left|{\rm det}_{n,k=1}^{\N} [\la\phi_{n}(0)|\phi_{k}(t)\ra]\right|^2.
\eeqa
Once the initial state is prepared, after the Zeno time scale,
each of the single-particle states
$\{|\phi_k\ra\}$ involved in the Slater determinant (\ref{2b}) experiences
a nearly exponential decay, i.e.
$|\phi_k(t)\ra\approx\exp(-\gamma_kt)||\phi_k(0)\ra$, $k=1,\dots,\N$.
Then, the $\N$-particle survival and non-escape probabilities decay exponentially,
 \begin{eqnarray}
\label{3ca}
\mathrm{P}_\N(t)&\approx& \exp(-\Gamma t)\mathcal{P}_\N(0),  \nonumber\\
\mathrm{S}_\N(t)&\approx& \exp(-\Gamma t)\mathcal{S}_\N(0),
\end{eqnarray}
with a decay rate
\begin{equation}
\label{4c}
\Gamma=2\sum_{k=1}^\N \gamma_k.
\end{equation}
Individual decay constants $\gamma_k$ are given by the imaginary parts of the complex energies
$E_k$ corresponding to the first $\N$ leading poles, $q_k$,
of the reflection amplitude for the potential trap  in the complex momentum $q-$plane,
$\gamma_k={\rm Im} E(q_k)= {\rm Re} q_k {\rm Im} q_k$.
One can avoid precise determination of the momentum pole positions by making instead a simple semiclassical estimate \cite{BZP}.
Let $E_k$ be the energy of the $k$-th bound state in the initial trap, and
 $x^k_0< x^k_1<x^k_2$ the three turning points in the quenched potential satisfying $V(x^k_i)=E_k$ ($i=0,1,2)$, e.g. for a potential where the left barrier is arbitrarily wide so that tunneling losses occur only through the right barrier, as in the case depicted in Fig. \ref{fcs}.
The semiclassical probability to tunnel in one attempt across the potential barrier is given by (see e.g, \cite{Bohm})
\begin{equation}
\label{4ca}
T(E_k)\approx \frac{\exp[-2S(E_k)]}{\{1+\exp[-2S(E_k)]/4\}^2}.
\end{equation}
where $S(E_k)\equiv \int_{x^k_1}^{x^k_2}\d x\{2[V(x)-E_k]\}^{1/2}$ is the complex
action corresponding to the classically forbidden region $x_1 < x < x_2$.
Since the period of the bound motion is given by
 $\tau_k = 2\int_{x_0^k}^{x_1^k}\d x\{2[V(x)-E_k]\}^{-1/2}$,
the particle impacts on the barrier with an approximate frequency $n_k=1/\tau_k$.
Multiplying $T$ by the number of impacts per unit time gives individual decay rates
\begin{equation}
\label{5c}
2\gamma_k\approx \tau_k^{-1}T(E_k),\q k=1,\dots,\N,
\end{equation}
which together with Eq. (\ref{4c}) yield the desired rate of the exponential decay in Eqs. (\ref{3ca}).

%

\section{The model and results \label{secmodel}}
%
 \begin{figure}
\includegraphics[width=0.8\linewidth,angle=0]{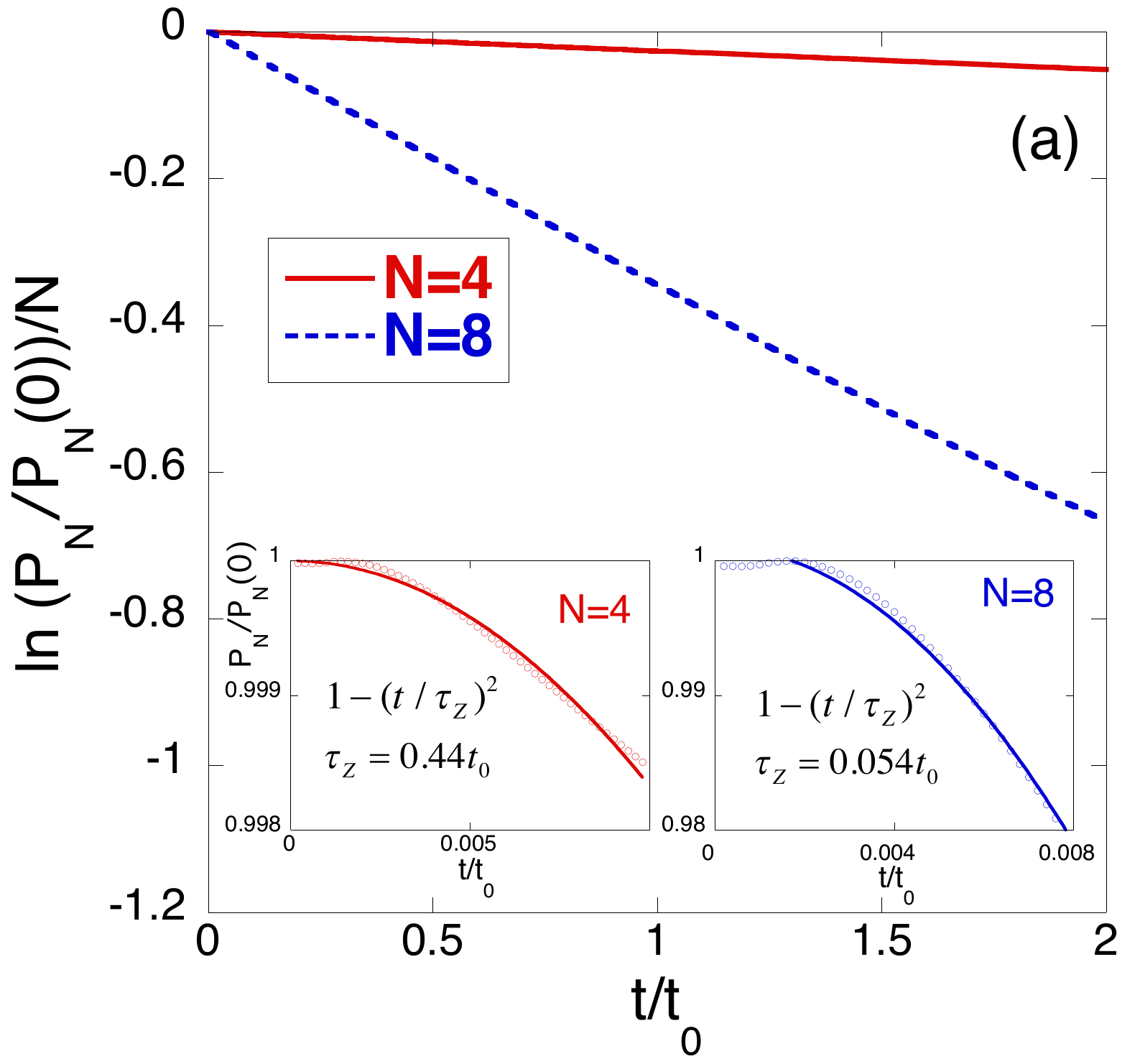}
\includegraphics[width=0.8\linewidth,angle=0]{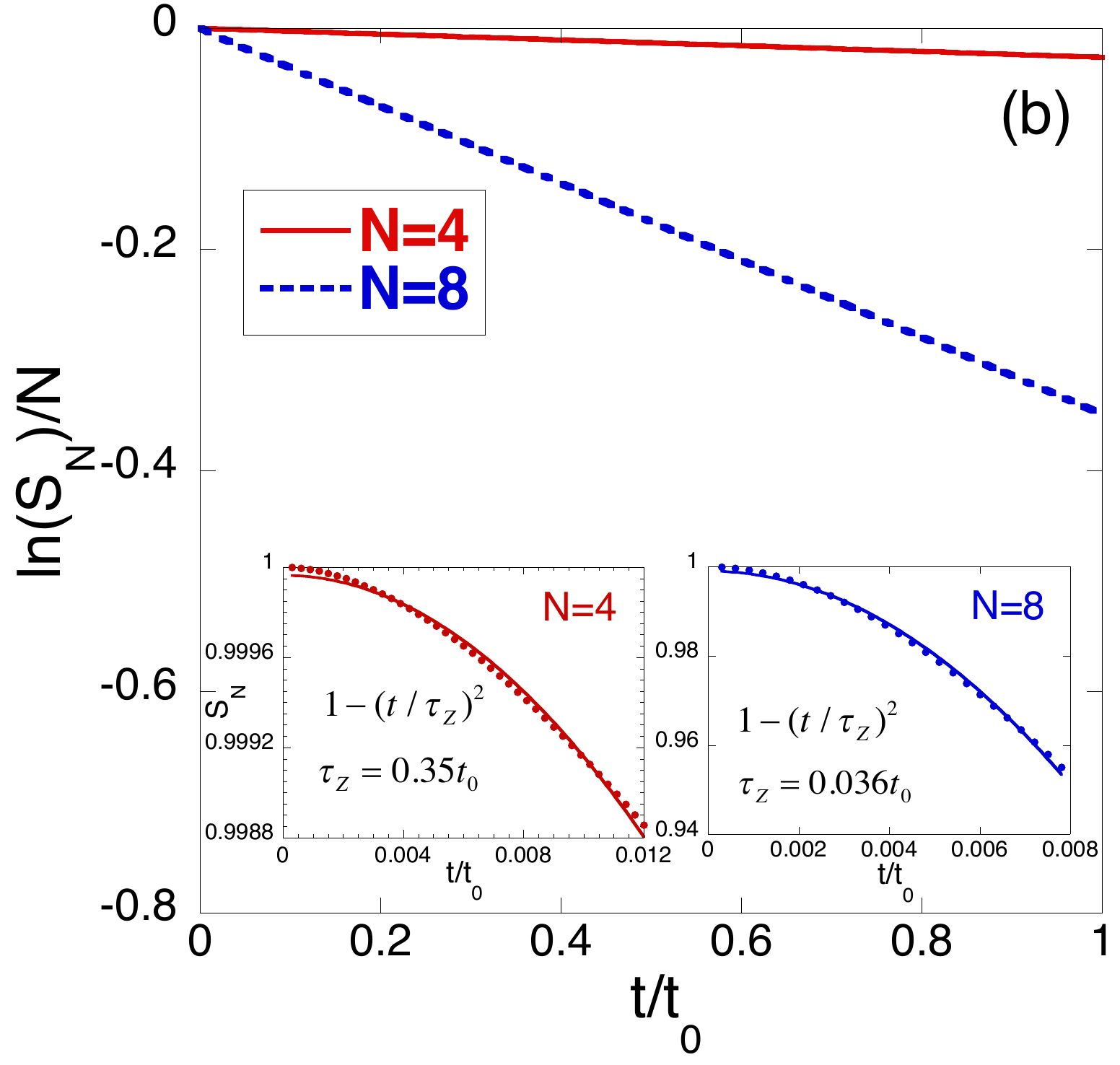}
\caption{Non-escape and survival probabilities for fermionic atom-number states with $\N=4$ and $\N=8$ leaking out of  potential (\ref{2d}) with $C=8$. The insets show the quadratic behavior of  $\mathrm{P}_{\rm N}(t)$ and $\mathrm{S}_{\rm N}(t)$ in the short-time regime (solid) and the fitted quadratic parabola (dots).}
\label{PSHCI}
\end{figure}
%
As a realistic model of a one-dimensional trap we consider a smooth bathtub potential (Fig. \ref{fcs}, dashed)
\beqa\label{1d}
\label{initialpot}
V_I(x)=-\frac{1}{2}V_0\bigg[1-\tanh\left(\frac{|x|-L/2}{\sigma_1}\right)\bigg]
\eeqa
whose right wall is instantly turned into a barrier of finite width at $t=0$ (Fig. \ref{fcs}, solid),
\beqa
\label{finalpot} \label{2d}
&&V(x,t)=-\frac{1}{2}V_0\bigg[1-\tanh\left(\frac { |x| -L/2}{\sigma}\right)\bigg]\Theta(a-x) \nonumber\\
&&-\frac{1}{2}V_0\bigg[1+\tanh \left(\frac { x-a -L_1/2}{\sigma_1}\right)\bigg]\Theta(x-a).\nonumber\\
\eeqa
Here $V_0$ is the depth of the initial well (and also the barrier height of the metastable potential),
$L$ and $L_1$ are the widths of the well and barrier, respectively, and $\sigma$ ($\sigma_1$)
determine the smoothness of the inner (outer) potential walls.
It is convenient to introduce the dimensionless variables
(we re-introduce the Plank's constant $\hbar$)
\beqa\label{3d}
x\to x/L, \q t\to t/t_0,\q
 V_0\to V_0t_0/ \hbar\q\
\eeqa
with $t_0\equiv mL^2/\hbar$.
%
 \begin{figure}
 \includegraphics[width=0.8\linewidth,angle=0]{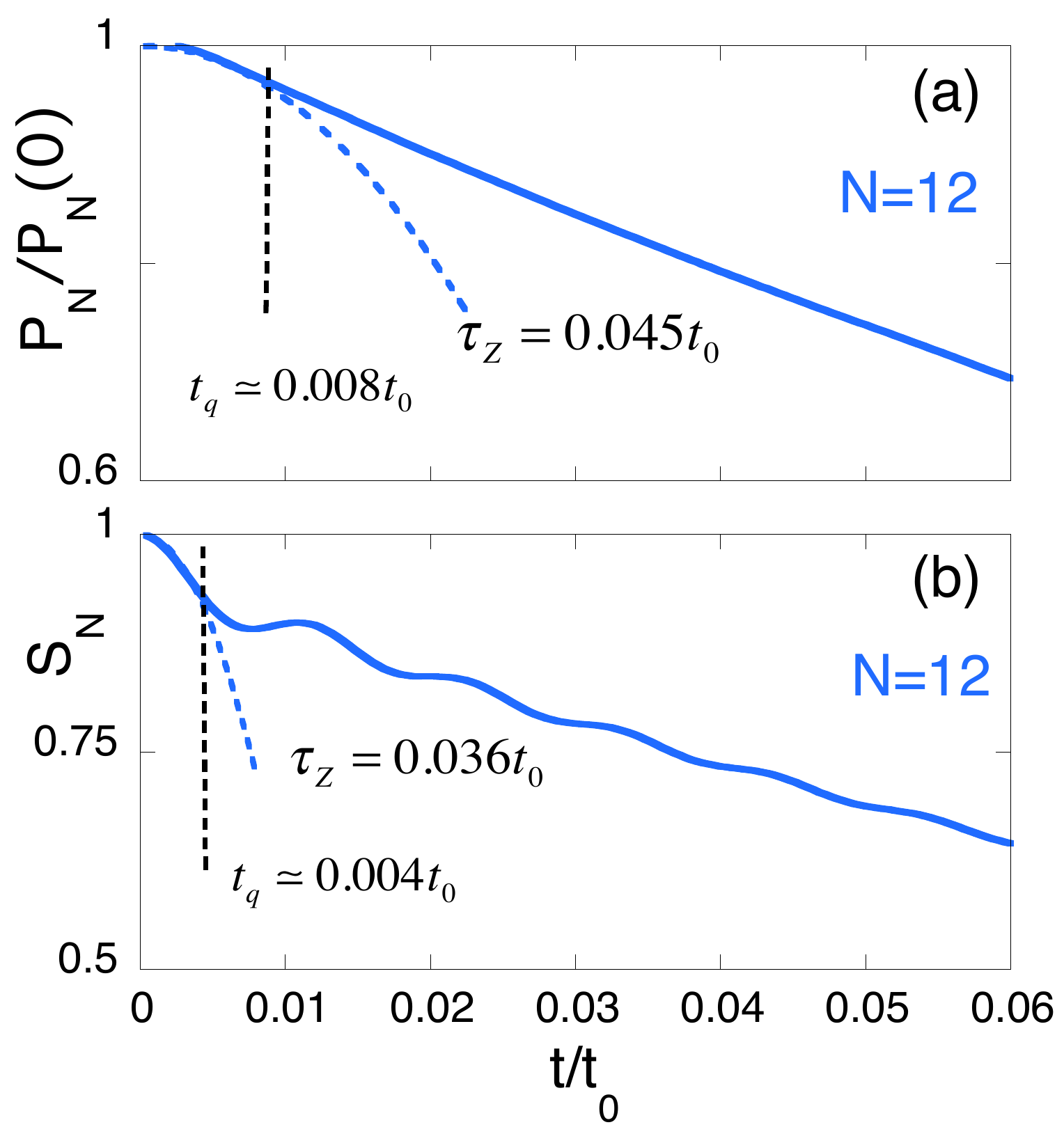}
\caption{Non-escape (a) and survival (b) probabilities for a fermionic 12-atom state leaking out of  potential trap in Eq. (\ref{2d}) with capacity $C=12$ (solid) (all times are in units of $t_0$). Also shown are the short time quadratic fits (dashed) and the times $t_q$ beyond which the quadratic approximation fails (vertical dashed).}
\label{PSdecay}
\end{figure}
%
The absorbing potential introduced by Manolopoulos \cite{Manolo,MugaPalao} is employed to avoid unphysical reflections at the boundaries of the numerical grid, $L^{box}_{1,2}$.
Bound one-particle eigenstates of the initial well, $\phi_n(x,0)$, are obtained by a standard finite-difference technique, and then evolved in time using the Crank-Nicholson scheme,
to yield $\phi_n(x,t)$ required in Eqs. (\ref{2c}) and (\ref{3c}). We use $L_1/L=0.08$, $\sigma/L=\sigma_1/L=0.01$, $a/L=0.55$ and $V_0t_0/\hbar=C^2\pi^2$, where $C$ is the capacity of the well, i.e., the maximum number of bound states it supports. We also note that for $^{23}$Na atoms in a potential well
with $L=80 \q \mu $m $t_0$ is about $2.39$ s. Finally, we chose $L^{box}_{1}/L=-20$, $L^{box}_{2}/L=30$, and the absorbing potential identical to that used in \cite{DIMA}.
For the calculation of the non-escape probability (\ref{2}) for a fermionized system in its ground state (\ref{2b}) we chose the spatial region $\Delta$ in Eq. (\ref{1c}) to include most of the trap,
\beqa\label{4d}
\Delta = (-\infty, a],
\eeqa
so that $\delta \varrho_0$ in Eq. (\ref{4}), now associated with exponential tails of the $\phi_n(x)$
extending into the right classically forbidden region of $V_I(x)$, is small.
Figure \ref{PSHCI} shows the decay dynamics of $\mathrm{P}_\N$ and $\mathrm{S}_\N$ for different fermionic atom-number states when the initial well supports a maximum of $C=8$ bound states. The two curves are remarkably similar, given that $\mathrm{S}_\N$ is sensitive to the population of individual one-particle states, while $\mathrm{P}_\N$ only depends on whether atoms have left the region $\Delta$.
In agreement with the discussion in the previous sections, two regimes can be identified.
The short-time behavior is characterized by a decay quadratic in time as shown in the inset of Fig. \ref{PSHCI}.
Subsequently, as dictated by Eq. (\ref{3ca}), the exponential law holds. The associated transition is illustrated in Fig. \ref{PSdecay} for a $N=C=12$ fermionic Fock state, and occurs approximately in the time scale
\beqa
\tau_q\approx\tau_Z^2\Gamma,
\eeqa
where the Zeno time $\tau_Z$ and decay rate $\Gamma$ are given by equations (\ref{5b}) and (\ref{4c}), respectively.

%
\begin{figure}
\includegraphics[width=0.8\linewidth,angle=0]{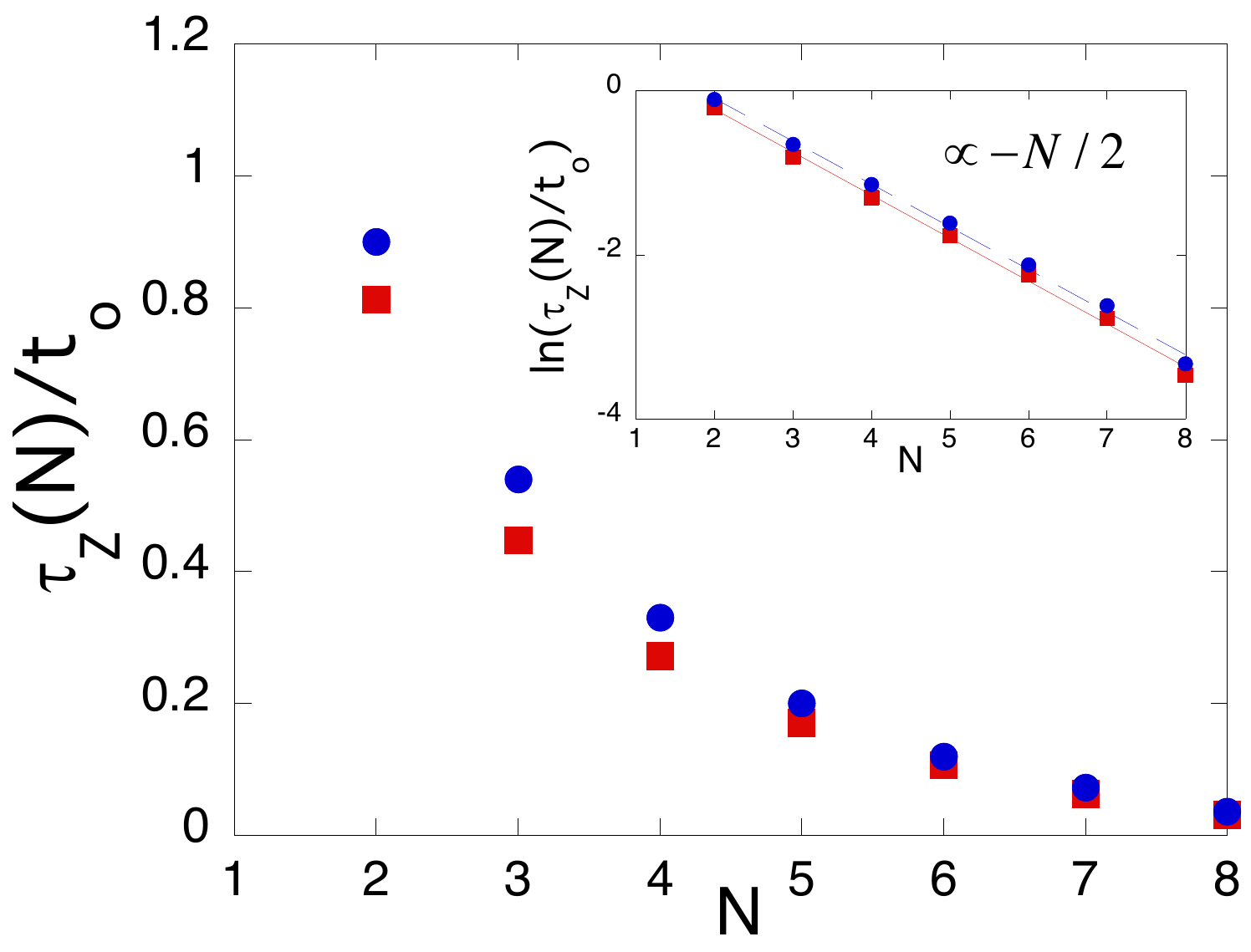}
\caption{Dependence of the survival Zeno time $\tau_Z$ on the number of trapped fermions $\N$ for
a potential (\ref{2d}) with $C=8$. The analytical Zeno times given by Eq. (\ref{5b}) (squares)
are shown together with the numerical values extracted from a parabolic fit to the exact short-time decay dynamics  (dots). The inset shows the corresponding exponential fits to the dependence of $\tau_Z$ on $\N$.}
\label{QZT}
\end{figure}
%
We first focus on the short-time dynamics, where the characteristic scale of the decay is given by the Zeno time $\tau_Z$. As the energy dispersion of the initial state with respect to the Hamiltonian for $t>0$ increases, the corresponding $\tau_Z$ is reduced. In particular, this is expected from Eq. \ref{5b} for increasing particle number $\N$. Figure \ref{QZT} shows the survival Zeno time $\tau_Z$ as a function of $\N$.  Its dependence on the particle number is enhanced as $N$ approaches $C$ and the initial Fock state involves energy components closer to the brim of the trap, which result in the spatial extension of the atomic cloud beyond the trap region $\Delta]$ due to existence of tunneling tails for $x>a$. Indeed, it is found numerically that
 the suppression of the Zeno-time is approximately exponential with respect to $\N$, $\tau_Z(\N)\approx \tau_Z(1)\exp(-\N/2)$, see the inset in Fig. {\ref{QZT}}.

%
 \begin{figure}
 \includegraphics[width=0.8\linewidth,angle=0]{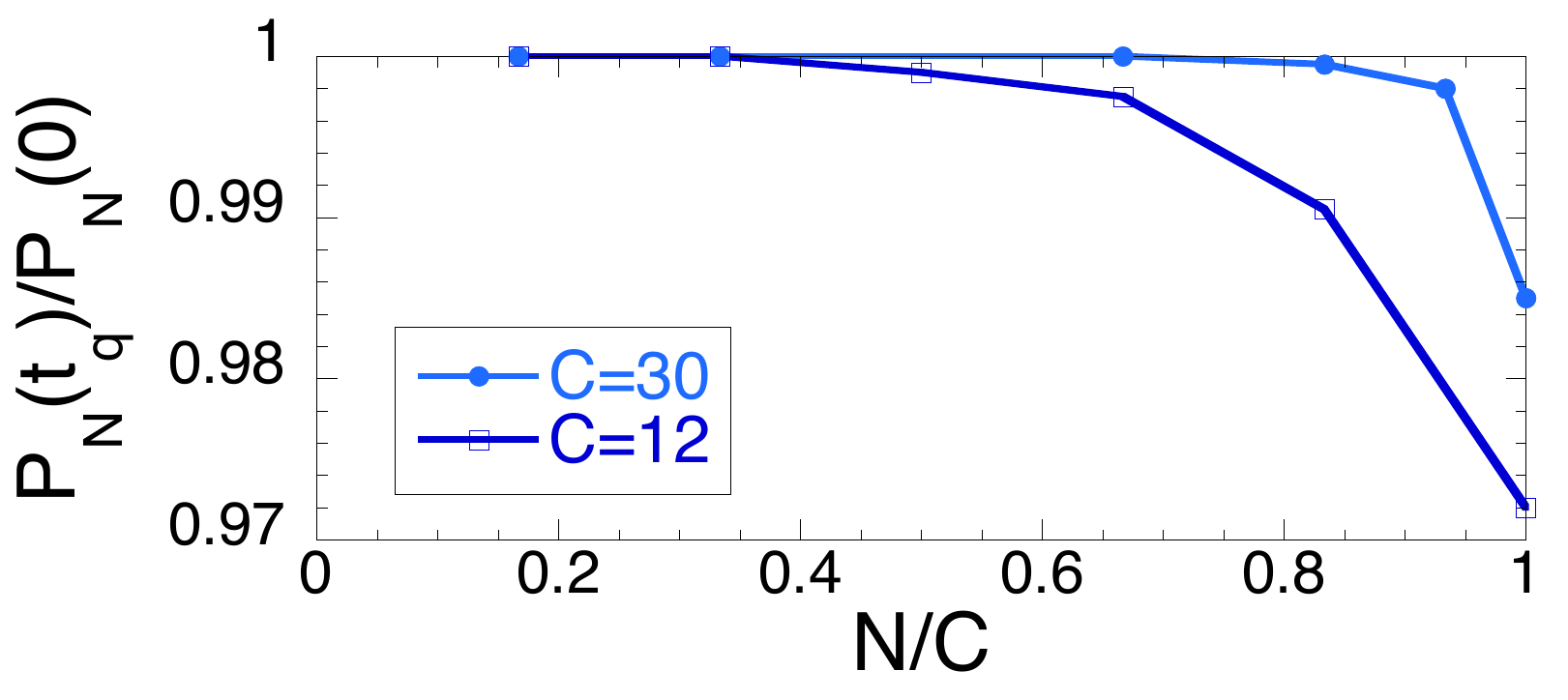}
\caption{Loss of fidelity for fermionic atom-number states  leaking out of  potential trap described by Eq. (\ref{2d}) for two different capacities $C=12,30$ ($\N\leq C$). The fidelity is evaluated at the transition time $t_q$  beyond which the quadratic approximation fails (cf. Fig. \ref{PSdecay}), this is, right before the exponential regime sets in.}
\label{PSfinal}
\end{figure}
This rapid decline of $\tau_Z(\N)$ might suggest that the short-time deviations from the exponential law should be irrelevant for Fock states with already moderate numbers of atoms. However, this is not completely the case, as is illustrated in Fig. \ref{PSdecay} which demonstrates that the fidelity decay of an Fock state also accelerates with the number of atoms, and becomes significant already at the end of the quadratic evolution, provided $\N$ is sufficiently close to $C$. 
To quantify this effect, we evaluated the transition time $t_q(\N)$ (Fig. \ref{PSdecay}, vertical dashed) such that for $t \gtrsim t_q$ the quadratic approximation to $\mathrm{P}_\N(t)$ or $\mathrm{S}_\N(t)$ starts to fail. (Note that $t_q$ is not the same as $\tau_Z$, which determines the curvature of the short-time parabolic decay.) 
The plot of the  fidelity factor $\mathrm{P}_\N(t_q)/\mathrm{P}_\N(0)$ vs. $\N$ shown in Fig. \ref{PSfinal} confirms that as $\N/C$ approaches unity a significant fidelity decay of an atom number state occurs before the exponential regime sets in. The effect is more pronounced for small traps, the goal of current experimental efforts.
\begin{figure}
\includegraphics[width=0.8\linewidth,angle=0]{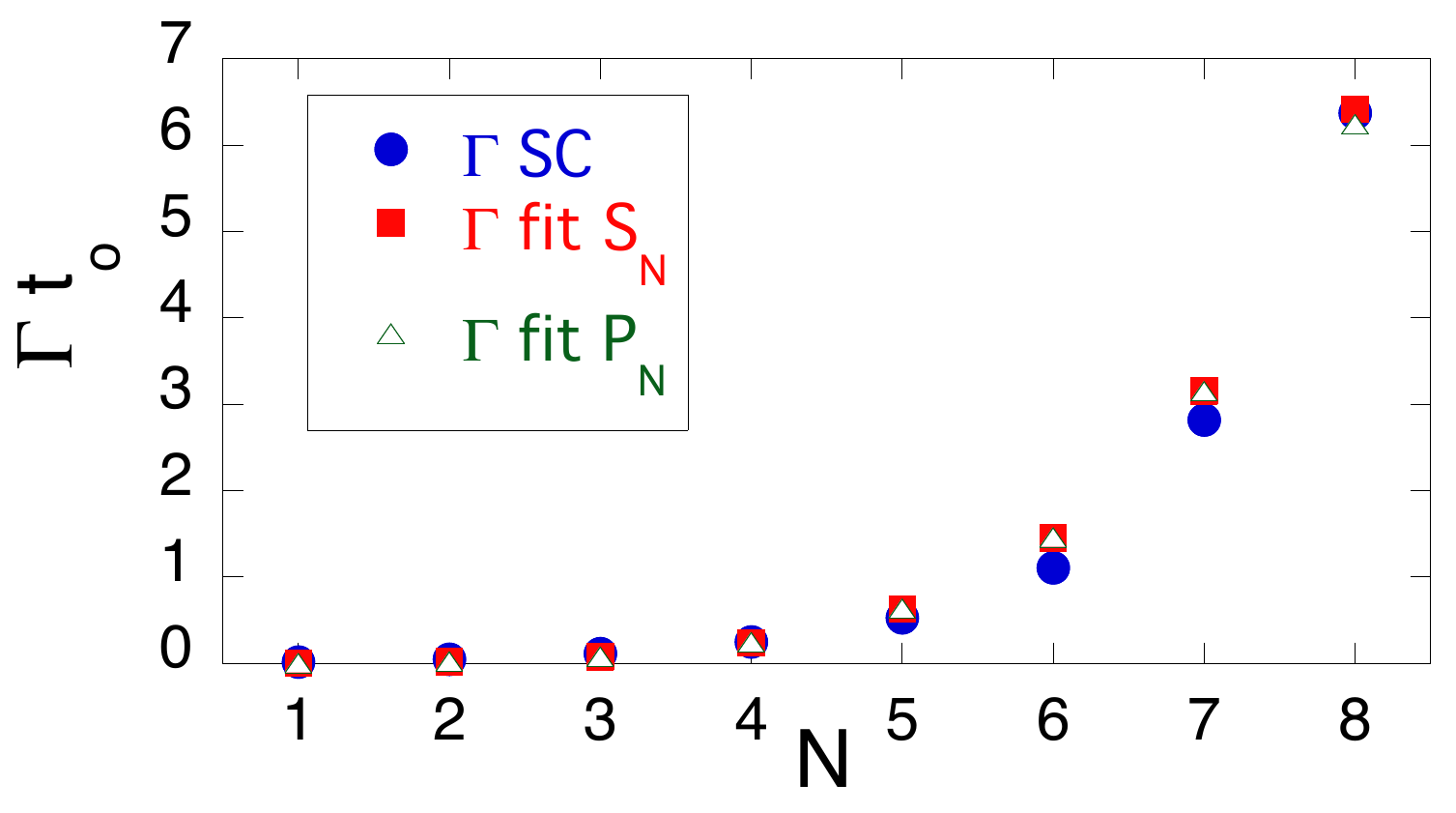}
\caption{Dependence of the exponential decay rates for  $\mathrm{S}_{\rm N}(t)$ (square) and $\mathrm{P}_{\rm N}(t)$ (triangle) on the number of trapped atoms $\N$ for a potential trap in Eq. (\ref{2d}) of capacity $C=8$. The observed survival decay rates are in excellent agrement with the analytical prediction based on Eqs. (\ref{5c}).}
\label{figrates}
\end{figure}

We next focus on the characterization of the exponential regime.
The dependence of the decay rate (\ref{4c}) on the number of particles $\N$ is in good agreement with the semiclassical estimate derived from Eqs. (\ref{5c}), as shown in Fig. \ref{figrates}. 
This is to be contrasted  with the case of a trap with thin potential barriers \cite{DDGCMR06} where the decay rate of any single-particle excited state is soon governed by the longest-lived resonance, favoring the observation of multiple exponential regimes (for $\N=2$ with a $\delta$-barrier, see \cite{GCML11}). In such case the multiparticle decay rate becomes $\Gamma=2\gamma_1\N$ linear in $\N$. 
Note that  the expression Eqs. (\ref{4c}) for the decay rates equally applies to distinguishable particles or excited states of non-interacting bosons.
Its success in reproducing the decay rates extracted from exponential fits to the exact decay dynamics, points out that the indistinguishability of the particles and the many-body correlations among them play a subdominant role in the exponential regime. 
Nonetheless, as the initial state includes states closer to the brim of the trap, Eq. (\ref{4ca}) losses accuracy.

\section{Discussion\label{discuss}}

Given the recent outburst of works dealing with few-body quantum decay \cite{DDGCMR06,Ceder09,delcampo11,Goold11,Buljan11,MG10,TS11,GCML11,KB11,KLW11}, it is interesting to establish the current understanding and point out some aspects deserving further studies. In particular, combining the results obtained in this manuscript with those in \cite{delcampo11}, we can identified the following stages in multi-particle quantum decay of systems with contact-interactions.
As in the single-particle decay, three regimes are found:

\begin{itemize}
\item {\it Short time asymptotics: Zeno regime. -}
A quadratic-in-time quantum decay occurs that is characterized by the multi-particle Zeno time, i.e. the inverse of the energy variance of the quenched Hamiltonian (that, for $t>0$) evaluated in the initial state, Eq. (\ref{5b}). The main contribution to this time scale arises from the energy spread of the initial state. Nonetheless, in the presence of interactions there is a counterintuitive correction reflecting the indistinguishability of the particles. This correction slows down the decay of polarized fermions with respect to degenerated states of non-interacting bosons. It would be clarifying to consider systems with Generalized Exclusion Statistics \cite{Haldane91} and compare the multi-particle quantum decay between systems with finite-interactions related by Bose-Fermi duality \cite{GNO04}.

\item {\it Exponential regime. - }
Following the Zeno regime, exponential decay sets in, characterized by multi-particle decay rates well described in a semi-classical approximation, i.e. Eq. (\ref{5c}). The dependence of the decay rates
is dominated by the energy spread of the initial state.
This is expected to be the general behavior for smooth potentials, but for finite-interactions the question
remains as to the validity of the semi-classical approximation. For example, considering a 1D Bose gas \cite{KB11}, does Eq. (\ref{5c}) evaluated at the quasi-momenta of the initial state (Bethe roots) describe accurately the decay rate, or are there important corrections arising from correlations among the particles?
Interestingly enough, the exact decay dynamics of two fermions for arbitrarily thin potential barriers
exhibits a transition among different decay rates \cite{GCML11}. The observability of multiple exponential regimes for arbitrary particle number $\N$ and as a function of the potential features well deserves further studies.

\item {\it Long-time asymptotics: post-exponential power-law decay. -}
The subsequent regime is characterized by a power-law decay and arises due to the possibility that the decay products recombine to form the initial state in a classical sense.
This regime is governed by the short-range density-density correlations, that lead to a dramatic dependence of the power-law exponents on the particle-number $\N$ \cite{delcampo11}. The current understanding would benefit from studies for systems with finite and long-range interparticle interactions as well as two and three dimensional systems with degeneracy arising from angular momentum. The power-law exponents in these systems are expected to exhibit a rich dependence on $\N$.

\end{itemize}

We note that a deep understanding of multi-particle quantum decay is relevant across very different fields of physics, beyond quantum foundations. For instance, a good control of the particle-number might be key to advance the field of quantum simulation with ultracold gases and this goal is being currently pursued in different laboratories \cite{exp1,exp2}. Similarly,  the short-time quantum decay is at the heart of certain applications such as dynamical decoupling schemes for high-fidelity quantum memories, quantum gates, quantum computing and generally to extend the coherence times in quantum systems \cite{FLP04}.
And at the same time, the existence of different stages of quantum decay may have important cosmological implications \cite{KD08}.

\section{Conclusions  \label{conclu}}

In summary, we have analyzed the many-body tunneling decay of trapped fermionic atom-number (Fock) states in a realistic model of relevance to recent experiments \cite{exp1,exp2}.
We focused our attention on the fidelity decay resulting from tunneling leakage, quantified as a decrease in the non-escape or survival probability, and identified the signatures of contact interactions and quantum statistics in the short-time multi-particle decay.
Even though the survival and non-escape Zeno times rapidly decrease with the number of atoms $\N$, as $\N$ approaches the maximum capacity of the trap a substantial loss of fidelity is found already at the end of the quadratic (Zeno) evolution, before the exponential regime sets in. Moreover, explicit expressions for the decay rates in the exponential regime have been provided.
Our results are amenable to experimental verification by standard techniques, e.g.,  by registering the time-evolution of the full-counting statistics
in a trap which allows leaking by quantum tunneling (see, for example, Ref. \cite{exp1}).
Finally, we note that strong $s$-wave scattering in ultracold bosons or spin-polarization in fermions suppress three body-losses, facilitating the study of genuinely quantum losses in these systems.

\section{Acknowledgement}

The authors are indebted to I. L. Egusquiza, G. Garc\'ia-Calder\'on, M. D. Girardeau and M. G. Raizen for valuable discussions and a careful reading of the manuscript. We acknowledge support of University of Basque Country UPV-EHU
(Grant GIU07/40), Basque Government (IT-472-10), and  Ministry of
Science and Innovation of Spain (FIS2009-12773-C02-01 and FIS2008-01236).


\begin{thebibliography}{10}
\bibitem{DRN07} A. M. Dudarev, M. G. Raizen, and Q. Niu, Phys. Rev. Lett. {\bf 98}, 063001 (2007).
\bibitem{MUG} A. del Campo and J. G. Muga, Phys. Rev. A {\bf 78}, 023412 (2008).
\bibitem{PONS} M. Pons, A. del Campo, J. G. Muga, and M. G. Raizen, Phys. Rev. A {\bf 79}, 033629 (2009).
\bibitem{WRN09} S. Wan, M. G. Raizen, and Q. Niu, J. Phys. B {\bf 42}, 195506 (2009).
\bibitem{RWZN09} M. G. Raizen, S. P. Wan, C. Zhang, and Q. Niu, Phys. Rev. A {\bf 80}, 030302(R) (2009).
\bibitem{DIMA} D. Sokolovski, M. Pons, A. del Campo, and J. G. Muga, Phys. Rev. A {\bf 83}, 013402 (2011) .
\bibitem{Qdistill} F. Heidrich-Meisner, S. R. Manmana, M. Rigol, A. Muramatsu, A. E. Feiguin, and E. Dagotto, Phys. Rev. A {\bf 80}, 041603(R) (2009).
\bibitem{NP10}  G. M. Nikolopoulos and D. Petrosyan, J. Phys. B {\bf 43}, 131001 (2010).
\bibitem{exp1} C. -S. Chuu, F. Schreck, T. P. Meyrath, J. L. Hanssen, G. N. Price, and M. G. Raizen. Phys. Rev. Lett. {\bf 95}, 260403 (2005).

\bibitem{exp2} F. Serwane, G. Z\"urn, T. Lompe, T. B. Ottenstein, A. N. Wenz, S. Jochim,
Science {\bf 332}, 336 (2011).

\bibitem{DDGCMR06} A. del Campo, F. Delgado, G. Garc\'ia-Calder\'on, J. G. Muga, and M. G. Raizen, Phys. Rev. A {\bf 74}, 013605 (2006).
\bibitem{Ceder09} A. U. J. Lode, A. I. Streltsov, O. E. Alon, H.-D. Meyer,
and L. S. Cederbaum, J. Phys. B: At. Mol. Opt. Phys. {\bf 42},  044018 (2009); {\bf 43}, 029802 (2010).
\bibitem{Girardeau60} M. D. Girardeau, J. Math. Phys. {\bf 1}, 516 (1960).
\bibitem{g3bosons} V. V. Cheianov, H. Smith, and M. B. Zvonarev, Phys. Rev. A {\bf 73}, 051604(R) (2006).
\bibitem{g3fermions} A. del Campo, J. G. Muga, and M. D. Girardeau, Phys. Rev. A {\bf 76}, 013615 (2007).
\bibitem{delcampo11} A. del Campo, Phys. Rev. A {\bf 84}, 012113 (2011).
\bibitem{Goold11} J. Goold, T. Fogarty, N. LoGullo, M. Paternostro, and T. Busch, Phys. Rev. A {\bf 84}, 063632 (2011).
\bibitem{Buljan11} K. Lelas, T. \v Seva, H. Buljan, Phys. Rev. A {\bf 84}, 063601 (2011).
\bibitem{MG10} A. Marchewka and E. Granot, arXiv:1009.3617.
\bibitem{TS11} T. Taniguchi and S. I. Sawada, Phys. Rev. E {\bf 83}, 026208 (2011).
\bibitem{GCML11} G.  Garc\'ia-Calder\'on, and L. G. Mendoza-Luna, Phys. Rev. A {\bf 84}, 032106 (2011).
\bibitem{KB11} S. Kim and J. Brand, J. Phys. B: At. Mol. Opt. Phys. {\bf 44}, 195301 (2011).
\bibitem{KLW11} A.  R. Kolovsky, J. Link, and S. Wimberger, arXiv:1112.6313.
\bibitem{Z1} P. Facchi, H. Nakazato, and S. Pascazio, Phys. Rev. Lett. {\bf 86}, 2699 (2001)
 \bibitem{Z2} P. Facchi and S. Pascazio, Phys. Rev. Lett. {\bf 89}, 080401 (2002);
J. Phys. A: Math. Theor. {\bf 41}, 493001 (2008).

\bibitem{shortexp} S. R. Wilkinson, C. F. Bharucha,
M. C. Fischer, K. W. Madison, P. R. Morrow,
Q. Niu, B. Sundaram, and M. G. Raizen, Nature {\bf 387}, 575 (1997).
\bibitem{SL} J. C. Slater,  Phys. Rev. {\bf 34}, 1293 (1929);
E. U. Condon,  Phys. Rev. {\bf 36}, 1121 (1930).
\bibitem{BZP} A. Baz, A. Perelomov, Ya. B. Zel'dovich, Scattering, {\it Reactions and Decay in Non- relativistic Quantum Mechanics}, (Israel Program for Scientific Translations, U.S. Dept. of Commerce, Clearinghouse for Federal Scientific and Technical Information, Springfield, 1969)
\bibitem{Bohm} D. Bohm,  {\it Quantum Theory}, (Dover, 1989)
\bibitem{Manolo} D. E. Manolopoulos, J. Chem. Phys. {\bf 117}, 9552 (2002).
\bibitem{MugaPalao} J. G. Muga, J. P. Palao, B. Navarro, and I. L. Egusquiza,  Phys. Rep. {\bf 395}, 357 (2004).
\bibitem{Haldane91} F. D. M. Haldane, Phys. Rev. Lett. {\bf 67}, 937 (1991).
\bibitem{GNO04} M. D. Girardeau, H. Nguyen, and M. Olshanii, Optics Communications {\bf 243}, 3 92004).
\bibitem{FLP04} P. Facchi, D. A. Lidar, and S. Pascazio, Phys. Rev. A {\bf 69}, 032314.
\bibitem{KD08} L. M. Krauss and J. Dent, Phys. Rev. Lett. {\bf 100}, 171301 (2008).


\end{thebibliography}
\end{document}